\documentclass[twocolumn,prl,nofootinbib]{revtex4}

\usepackage[caption=false]{subfig}

\usepackage{amsmath,amsfonts,amssymb,array}

\usepackage{color}
\usepackage{graphicx}

\def \be {\begin{enumerate}}
\def \ee {\end{enumerate}}
\def \beq {\begin{equation}}
\def \eeq {\end{equation}}
\def \ba {\begin{eqnarray}}
\def \ea {\end{eqnarray}}
\def \ban {\begin{eqnarray*}}
\def \ean {\end{eqnarray*}}

\def \bfl {\begin{flalign*}}
\def \efl {\end{flalign}}
\def \bsp {\begin{split}}

\def \l {\left}
\def \r {\right}

\newcommand{\ketbrad}[1]{|#1\rangle\!\langle #1|}

\newcommand{\mean}[1]{\langle#1\rangle}

\newcommand{\Mean}[1]{\left\langle#1\right\rangle}
\def\ket#1{\left| #1\right>}
\def\bra#1{\left< #1\right|}

\begin{document}
\title{Distinguishing Quantum and Classical Many-Body Systems}
\author{Dvir Kafri$^{1,2}$, Jacob Taylor$^{1,2,3}$}
\affiliation{1: Joint Quantum Institute, The University of Maryland, USA\\
2: Joint Center for Quantum Information and Computer Science, The University of Maryland, USA\\
3: National Institute of Standards and Technology, Gaithersburg, MD, USA}
\begin{abstract}
 Controllable systems relying on quantum behavior to simulate distinctly quantum models so far rely on increasingly challenging classical computing to verify their results. We develop a general protocol for confirming that an arbitrary many-body system, such as a quantum simulator, can entangle distant objects. The protocol verifies that distant qubits interacting separately with the system can become mutually entangled, and therefore serves as a local test that excitations of the system can create non-local quantum correlations. We derive an inequality analogous to Bell's inequality\cite{Bell1964,Clauser1969} which can only be violated through entanglement between distant sites of the many-body system. Although our protocol is applicable to general many-body systems, it requires finding system-dependent local operations to violate the inequality. A specific example in quantum magnetism is presented.
\end{abstract}
\maketitle

Quantum simulators can efficiently model quantum systems \cite{Feynman1982, Lloyd1996}. However, characterizing and validating such devices is in general difficult. Indeed, quantum state tomography\cite{Nielsen2000} for even eight qubits has required weeks of classical computational processing time\cite{Haffner2005} (in addition to exponentially growing measurement requirements). This issue is also present in process tomography, the analogue for quantum channels\cite{Chuang1997,Poyatos1997}. Notwithstanding the number of measurements growing exponentially in system size, for systems with $10+$ constituents reliable tomography is expected to break down from systematic errors in preparation and measurement\cite{Merkel2013}.  Although these resource scaling problems are partially alleviated with methods based on compressed sensing\cite{Gross2010,Gross2011,Shabani2011,Smith2012,Qi2013}, the cost still scales at least linearly with the Hilbert space dimension, and thus exponentially in the number of constituents. Added difficulties arise when data is limited to experimentally accessible local observables\cite{Kaznady2009}. 

Despite the costs of completely characterizing large quantum systems, there do exist scalable tests giving incomplete -- though useful -- descriptions of system behavior. For example, techniques such as randomized benchmarking\cite{Emerson2005,Knill2008,Wallman2014,Epstein2014} and fidelity estimation\cite{Silva2011,Flammia2011} require a number of measurements polynomial in system size, while still quantifying useful information such as error rates or average gate fidelities. In the case of locally correlated errors, this is sufficient to guarantee the operation of error-corrected quantum computer\cite{Gottesman1998,Aharonov1997, Preskill1998,Knill1998}. Recent linear optics experiments have likewise partially verified implementations of boson sampling, a task which currently cannot be carried out classically\cite{Aaronson2011}. Although it is impossible to efficiently verify this sampling, by using the results of Ref.~\cite{Aaronson2013} the authors of Ref.~\cite{Spagnolo2014} were able to distinguish the sampled distribution from a uniform one. The experiment of Ref.~\cite{Carolan2014} produced similar results, while also checking whether the photon statistics corresponded to indistinguishable particles. 

\begin{figure}
\centering
\includegraphics[width = 0.5\textwidth]{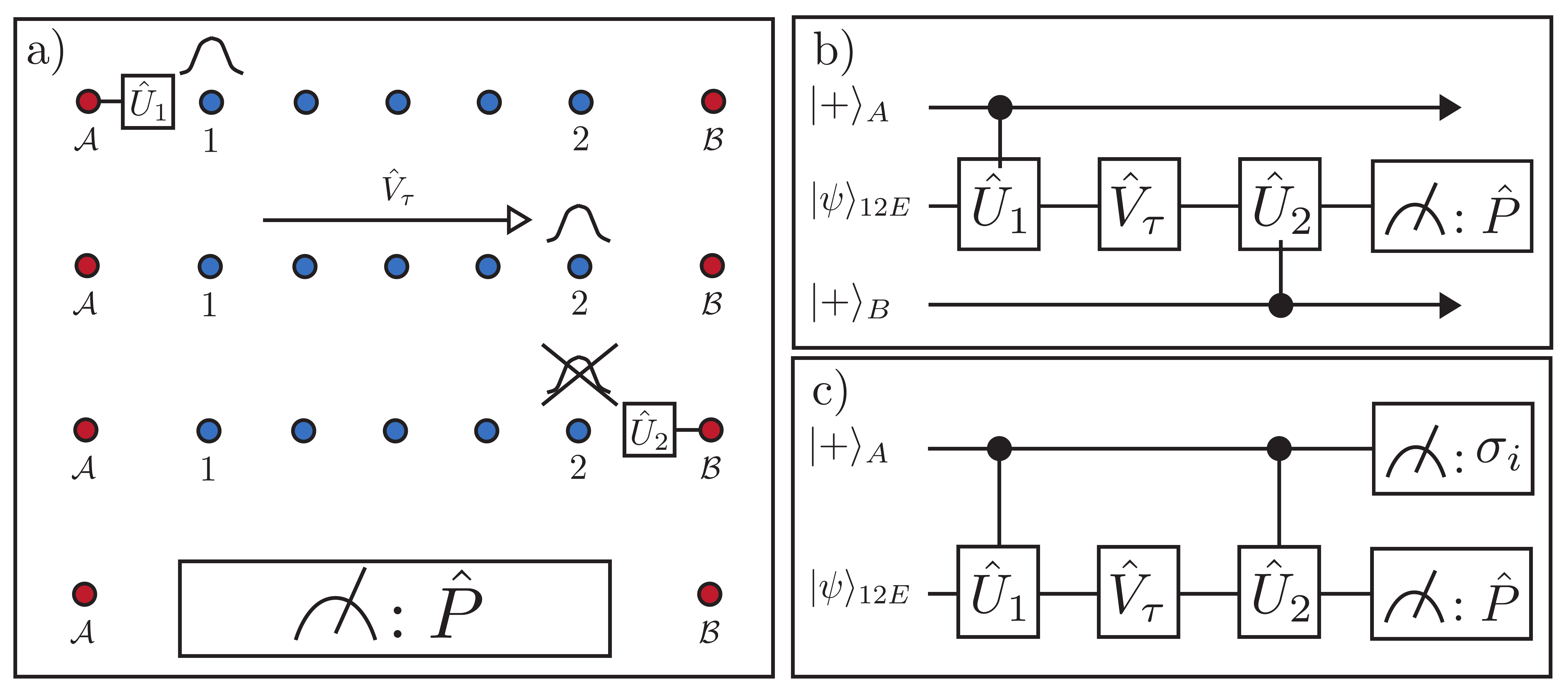}
\caption{{\bf a)} and {\bf b)}: Prototype entangling procedure. Ancilla qubit $A$ prepared as $\ket{+} = (\ket{0} + \ket{1})/\sqrt{2}$ applies controlled unitary $\hat U_1$ to site $1$ of the initial MBS state $\ket{\psi}_{12 E}$, producing a conditional excited state $(\ket{0}_A\otimes\ket{\psi}_{12E} + \ket{1} \otimes \hat U_1\ket{\psi}_{1 2 E})/\sqrt{2}$. Time evolution $\hat V_\tau$ propagates the MBS, spreading the excitation across the system. Ancilla qubit $B$ (also in superposition) then applies controlled unitary $\hat U_2$ to site $2$, removing the excitation (conditional on its initial state). Finally, a post-selection conditioned on projector $\hat P$ is carried out, confirming the MBS has returned to its original state. This projects the ancillas into a (possibly entangled) correlated state. {\bf c)} An analogous protocol using only one ancilla qubit. After post-selection, measurements of the single ancilla are sufficient to determine whether the previous protocol would have entangled the two ancillas.
}\label{circuit-diagram}

\end{figure}

This paper presents another incomplete test, in the context of quantum simulators of many-body systems (MBS)\cite{Buluta2009,Cirac2012}. It derives from a constructive procedure to entangle two distant ancilla qubits through local interactions with the many-body system. It starts by preparing the system in an initial, known state, composed of many spatially-fixed sites (see Fig.~\ref{circuit-diagram}). We apply a local perturbation to a single site, conditioned on the state of an ancilla qubit in superposition. The simulator then propagates the system forward in time, correlating the ancilla with other sites of the MBS. At one of these distant sites, we apply a second perturbation controlled by a second ancilla qubit. This interaction is chosen to increase the probability amplitude between the current excited MBS state and its initial state, but only does so if both qubits are in the same control state. Qualitatively, excitations induced by the first qubit are conditionally removed by the second. Although both ancillas are now correlated, their correlation with the MBS prevent them from being entangled. We therefore measure the MBS and post-select for it being in its original state. This both disentangles the MBS from the ancillas and increases the probability that they are in the same eigenstate. Finally, we can directly verify that the qubits are entangled (e.g., through state tomography).

As a toy model, we consider a one dimensional transverse field Ising Hamiltonian,
\beq
\label{ising}
\hat H = B \sum_i \sigma_x^{(i)} - J \sum_{i} \sigma_z^{(i)}\otimes \sigma_z^{(i+1)}\,.
\eeq
We write the evolution for a time $\tau$ as $\hat V_\tau = \exp(-i \tau \hat H)$. In the $B/J\ll 1$ limit, the eigenstates of $\hat H$ are well approximated by eigenstates of the transverse field term, $\hat H_J =-J \sum_i \sigma_z^{(i)}\otimes  \sigma_z^{(i+1)} $. We therefore assume an initial state of the form
\beq
\label{Bground}
\ket{g} = \ket{0}\otimes \ket{0}\otimes \,...\, \otimes \ket{0}\,,
\eeq
where $\sigma_z \ket{0} = \ket{0}$, which minimizes the energy $\Mean{\hat H}$ up to corrections of order $B \cdot(B/J)^N$. The lowest lying excitations of $\hat H_J$ are described by domain walls. For our protocol to produce entanglement, we use excitations that are indistinguishable from $\ket{g}$ outside a finite region. We therefore consider the evolution of the next lowest excitations of $\hat H_J$, involving pairs of domain walls of the form,
\ba
\label{2dwall}
\ket{e_{i, j}} = \prod_{i\leq k \leq j} \sigma_x^{(k)} \ket{g}\,,
\ea
where $1< i \leq j < n$. Evolution under $\hat H = \hat H_J + \hat H_B$ disperses these states across the chain\cite{Pfeuty1970}. 

We now consider how to use the MBS to entangle two distant qubits. First, an ancilla qubit (labeled $A$) in state $\ket{+} = \frac{1}{\sqrt{2}}(\ket{0} + \ket{1})$ rotates the {\it second}\footnote{Since the first spin only interacts with one neighbor, exciting it would create a single domain wall instead of two.} spin of the chain through controlled-NOT gate, $\hat U_{A,2} = \ketbrad{0}_A\otimes \hat 1 + \ketbrad{1}_A\otimes \sigma_x^{(2)}$. This produces the excitation $\ket{e_{2,2}}$ in the chain, conditioned on the state of the ancilla qubit,
\beq
\hat U_{A,2} \ket{+}_A \otimes \ket{g} =\frac{1}{\sqrt{2}}\l(\ket{0}_A \otimes \ket{g}  +  \ket{1}_A  \otimes \ket{e_{2,2}}  \r)\,.
\eeq
At this point, the ancilla state is correlated with the MBS, though only at spin 2 of the chain. 

Continuing with the entangling procedure, we propagate under $\hat V_\tau$ for a time $\tau \sim N/B$ in order to spread the correlation with the ancilla across the spin chain as a superposition of excitations. After applying $\hat V_\tau$, the $\ket{g}$ component of the MBS acquires only an irrelevant global phase $\theta $, while the excited state $\ket{e_{2,2}}$ disperses over the subspace spanned by $\{\ket{e_{i j}}\}$:
\beq
\label{step2}
\hat V_\tau \hat U_{A,2} \ket{+}_A \otimes \ket{g} = \frac{1}{\sqrt{2}}\l(e^{i \theta}\ket{0}_A \otimes \ket{g}  + \ket{1}_A \otimes \hat V_\tau \ket{e_{2,2}}  \r)\,.
\eeq
To understand this process, we consider the projection of $\hat H_B$ to the linear span of states $\ket{e_{i, j}}$, an approximation valid when $B/J\ll 1$\cite{Bravyi2011}. We note that $\hat H_B$ only couples between adjacent domain walls,
\beq
\label{CTQW}
\bra{e_{i, j}} \hat H_{B} \ket{e_{k, l}} = B (\delta_{i,k+1} + \delta_{i,k-1} + \delta_{j,l+1} + \delta_{j,l-1})\,,
\eeq
so the evolution of $\ket{ e_{2,2}}$ under $\hat H$ is equivalent to that of a continuous-time quantum walk in two dimensions\cite{Farhi1998b}. The distinct behaviors of $\ket{g}$ and $\ket{e_{2,2}}$ under $\hat V_\tau$ will allow us to distinguish the states of the ancilla through a local operation at spin $N-1$.

\begin{figure}
\centering
\includegraphics[width = .5\textwidth]{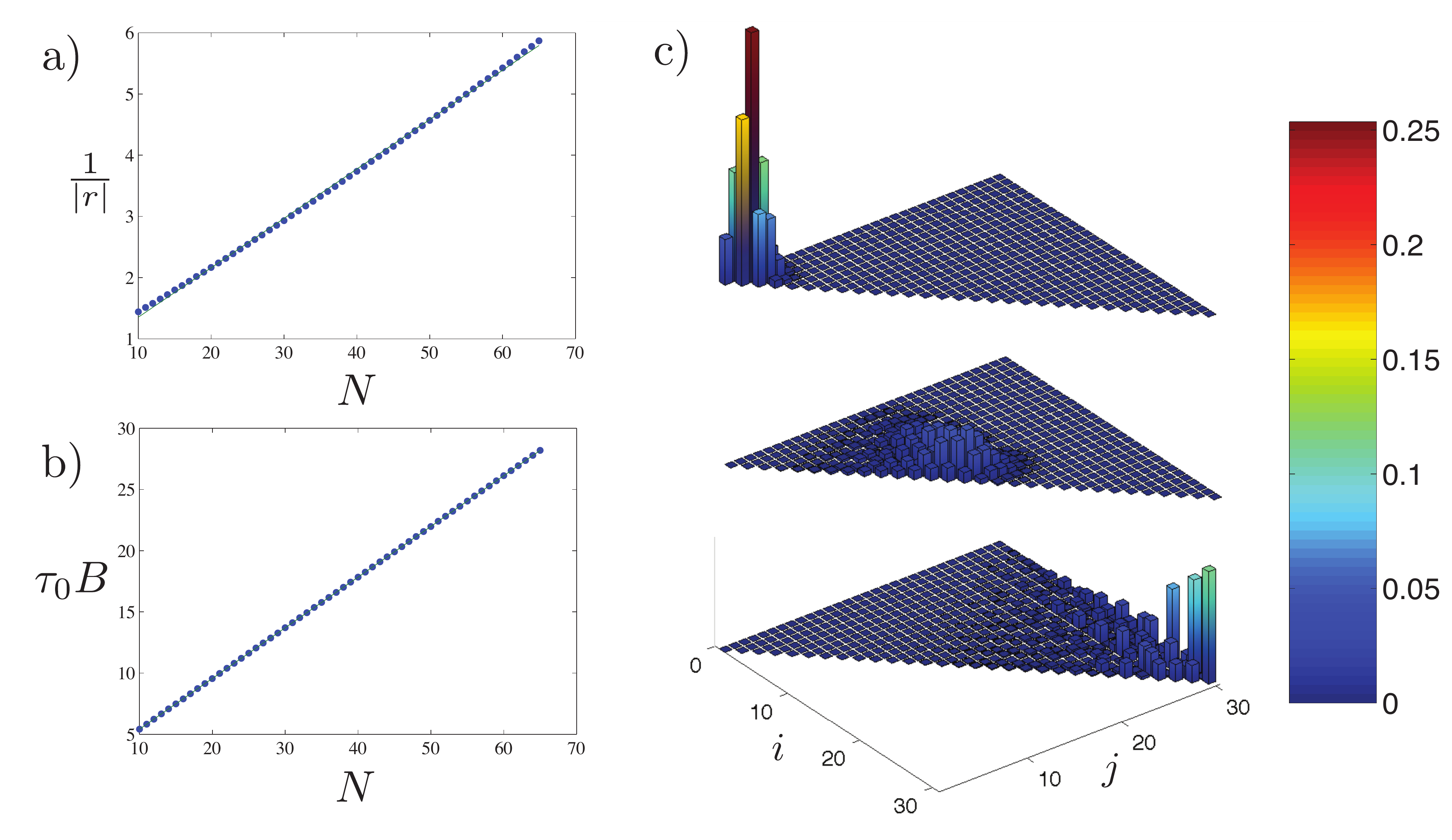}
\caption{Evolution time and ancilla qubit correlation as a function of spin chain length $N$, in the quantum walk approximation. {\bf a.)} Inverse plot of the peak values of $|r|  = |\bra{e_{N-1,N-1}} \hat V_\tau \ket{e_{2,2}}|$, as a function of chain length $N$. The numerically calculated values (dots) match closely to the linear fit, with $|r|^{-1} \simeq 0.08 N + 0.55$. {\bf b.)} A plot of the time $\tau_0$ taken for the overlap $|r|$ to reach its peak value. The numerically calculated values (dots) match the linear fit as $\tau_0 B \simeq 0.52 N + 2.02$. {\bf c.)} A simulation of state occupations $\Mean{\ketbrad{e_{i,j}}}$ in a quantum walk corresponding to $N = 32$ qubits. The system is initialized in state $\ket{e_{2,2}}$ at  time $\tau = 0$. The time lapse corresponds to times $\tau = 0.1 \tau_0$ (top), $\tau = 0.5 \tau_0$ (middle), and $\tau = \tau_0$, where $\tau_0$ corresponds to the peak time determined in the previous fit.  } \label{costCorrelation}
\end{figure}

After evolving for a time $\tau\sim N/B$, the excitation $\hat V_\tau \ket{e_{2,2}}$ has a probability $\sim 1/N^2$ of being localized in state $\ket{e_{N-1,N-1}}$. We verify this numerically by using the quantum walk analogy above. As seen in Fig. \ref{costCorrelation}a, the peak transition probability $|\bra{e_{N-1,N-1}} \hat V_\tau \ket{e_{2,2}}|^2$ scales as $\sim 1/N^2$. Key to the success of our protocol, we note that the unitary $\sigma_x$ at site $N-1$ maps the excitation $\ket{e_{N-1,N-1}}$ back to the ground state $\ket{g}$. Hence we can applying $\sigma_x^{(N-1)}$ to $\hat V_\tau \ket{e_{2,2}}$ to give it a non-zero overlap with the ground state, 
\beq
\label{overlap}
r = \bra{g} \sigma_x^{(N-1)}\hat V_\tau \ket{e_{2,2}} = \bra{e_{N-1,N-1}} \hat V_\tau \ket{e_{2,2}}\neq 0\,,
\eeq
where $|r|^2 \sim 1/N^2$. The time $\tau$ required for the overlap $|r|$ to reach its peak scales linearly with $N$ (Fig. \ref{costCorrelation}b), as opposed to a time scale $\sim N^2/B$ observed in diffusive propagation\cite{Kempe2003}.

We can use the fact that $\ket{e_{2,2}}$ can transition to $\ket{e_{N-1,N-1}}$ via $\sigma_x^{(N-1)}$ to entangle ancilla $A$ with a second ancilla, $B$. After time evolution $\hat V_\tau$, we apply a second controlled-NOT gate between $B$ and spin $N-1$,
\beq
\label{UBN1}
\hat U_{N-1, B} =  \ketbrad{0}_B\otimes \hat 1 + \ketbrad{1}_B\otimes \sigma_x^{(N-1)} \,.
\eeq
The resulting state displays correlations between both ancillas and the MBS,
\ba
\nonumber & &\hat U_{N-1,B} \hat V_\tau \hat U_{A,2} \ket{+\,+}_{AB}\otimes \ket{g}= \\ 
\nonumber && \frac{1}{2}\l(\ket{0\,0}_{AB} \otimes e^{i \theta} \ket{g}   + \ket{1\,0}_{AB} \otimes \hat V_\tau\ket{e_{2,2}}  + \r.\\
\nonumber && \l.  \ket{0\,1}_{AB} \otimes e^{i \theta} \sigma_x^{(N-1)}   \ket{g} +  \ket{1\,1}_{AB} \otimes \sigma_x^{(N-1)}  \hat V_\tau\ket{e_{2,2}}  \r) \,.
\ea
To motivate the final step, observe that if the ancillas are in either state $\ket{0\,1}_{AB}$ or $ \ket{1\,0}_{AB}$, the MBS is necessarily in an excited state. Thus if we measure and post-select the MBS to be in the ground state $\ket{g}$, we project the ancilla qubits into an entangled superposition of states $\ket{0\,0}_{AB}$ and $\ket{1\,1}_{AB}$. This post-selection succeeds with probability $\frac{1 + |r|^2}{4}$, and produces the entangled state, 
\ba
\bra{g} \hat U_{N-1,B} \hat V_\tau \hat U_{A,2} \ket{+\,+}_{AB}\otimes \ket{g} =\nonumber \\  \frac{1}{\sqrt{1 + |r|^2}}\l(e^{i\theta} \ket{0 \,0}_{AB} +  r \ket{1\,1}_{A B} \r)\,.\label{final}
\ea
To conclude the protocol, a measurement of the ancillas would then confirm that the ancillas are entangled.

This simple approach can be generalized to a protocol on generic many-body systems. We consider a propagator for an arbitrary many-body Hamiltonian $\hat V_\tau = \exp(-i \hat H \tau)$ (which may also represent environmental degrees of freedom) and an initial prepared ground state $\ket{\psi_{12E}}$ (or in the case of a mixed state environment, its purification over a larger subspace). The MBS is composed of two local components $S_1$ and $S_2$, with $E$ representing the rest of the system. As in the experimental example, we consider two ancilla qubits, interacting locally with $S_1$ and $S_2$ at different times. In full, the unitary evolution is
\beq
\label{fullevolution}
C_{B}(\hat U_2) \, \hat V_\tau \, C_{A}(\hat U_1) \ket{ +}_A  \ket{+}_B \ket{\psi_{12E}}\,,
\eeq
where $C_{A}(\hat U_{1}) =\ketbrad{0}_A \otimes \hat 1 + \ketbrad{1}_A\otimes \hat U_1$ (with $C_B(\hat U_2)$ defined analogously). Importantly, we assume that the local unitaries $\hat U_1,\hat U_2$ individually bring the MBS to an excited state, though the combined evolution $\hat U_2 \hat V_\tau \hat U_1 $ produces a non-zero overlap with the ground state. After the second controlled unitary, we make a post-selective measurement on the MBS represented by the projection operator $\hat P$. Although this projection can be arbitrary, we assume it confirms that the system has returned to its ground state. As we argued in the example case, this post-selection imparts a known correlation to the ancilla qubits.  

The actual test of the MBS derives from verifying that the ancilla qubits are entangled. Writing out the qubit density matrix in the $\sigma_z$ basis, we have 
\beq
\label{rho}
\rho_{i j, i' j'} = \frac{1}{p}\Mean{\hat (\hat U_1^\dagger)^{i'} \, \hat V_\tau^\dagger \,(\hat U_2)^{j'} \, \hat P \,(\hat U_2^\dagger)^{ j} \, \hat V_\tau \,(\hat U_1)^{i} }\,,
\eeq
where $\mean{\hat O} = \bra{\psi_{12E}} \hat O \ket{\psi_{12E}}$ refers to an average over the MBS state alone, and $p$ is the probability of making the projective measurement $\hat P$ for the state of Equation \eqref{fullevolution}. The product $(\hat U_2)^{ j} \, \hat V_\tau \,(\hat U_1)^{i}$ (with $i,j$ either $0$ or $1$) represents the unitary evolution of the MBS conditioned on the ancillas being in initial state $\ket{i j}$. Hence to generically determine whether the qubits are entangled, one may carry out full state tomography of the qubits' density matrix which can be done through concurrent local measurements on the individual qubits.

The protocol outlined above is the central result of our paper. Given the ability to prepare and measure an initial state, it provides a local test verifying that the MBS propagator can generate non-local entanglement. Such a result precludes a description in which subsystems are locally quantum but all correlations between subsystems are essentially classical. In this light, the protocol is akin to a Bell's inequality applied to the system and its dynamics as a whole. Like Bell's inequality, it uses only local operations. It also has a `loophole': we require that the sites of the MBS are spatially stationary. Otherwise a single site could migrate through the MBS and interact with both ancilla to entangle them, bypassing the need for quantum information to be passed between different sites of the MBS.

Although the procedure we have presented generically requires two ancilla qubits to be carried out, under certain assumptions this requirement may be loosened. Indeed, although full state tomography on the ancillas is required to generically verify entanglement, we note that by equation \eqref{rho} the qubit density matrix is completely determined by averages over the MBS alone. This means that in certain cases, even in the absence of one of the ancilla qubits, it is possible to test whether entanglement \textit{would} have occurred. Such a test derives from the Peres-Horodecki criterion\cite{Peres1996,Horodecki1996}, which states that the ancillas are entangled if and only if the partial transpose matrix $\rho^\Gamma$ has a negative eigenvalue.  This property is characterized by Sylvester's criterion, which states that a square matrix $A$ has no negative eigenvalues if and only if its principal minors are all non-negative\cite{Horn2012},\footnote{The principal minors of a square matrix $A_{m n}$ are the determinants $\det(A^{(s_1,s_2,\,...\, s_k)})$, where $A^{(s_1,s_2,\,...\,,s_k)}$ is the matrix $A$ truncated to only rows and columns $\{s_1,s_2,\,...\,, s_k\}$.}. Hence to confirm that the ancilla qubits are entangled, it is sufficient to check that a single principal minor of $\rho^\Gamma$ is negative: $\rho^\Gamma_{0 1,0 1}\rho^\Gamma_{1 0,1 0}  - \rho^\Gamma_{0 1, 1 0}   \rho^\Gamma_{1 0,0 1}  <0$. We map this statement to an expression on the MBS by using $\rho_{i j, i' j'}^\Gamma = \rho_{i j', i' j}$ and equation \eqref{rho}:
\beq
\label{ineq}
\Mean{\hat V^\dagger \hat U_2^\dagger \hat P \hat U_2 \hat V} \Mean{\hat U_1^\dagger \hat V^\dagger \hat P \hat V \hat U_1} <
\l|\Mean{\hat V^\dagger \hat P \hat U_2 \hat V \hat U_1}\r|^2\,.
\eeq
When this inequality holds, the ancilla qubits become entangled under our protocol. It can only be satisfied when quantum correlations are propagated between spatially distant sites. Importantly, since $\mean{\hat O} = \bra{\psi_{1 2 E}} \hat O \ket{ \psi_{1 2 E}}$ represents averages over only the state $\ket{\psi_{1 2 E}}$, so it characterizes the many-body system and its evolution alone.

Although inequality \eqref{ineq} implies the pair of qubits in the protocol become entangled, it can actually be measured using a single ancilla qubit. First, we note that the product of terms on the left hand side require no ancilla qubits to be measured: for example, the quantity $\mean{\hat V^\dagger \hat U_2^\dagger \hat P \hat U_2 \hat V}$ is simply the probability of measuring the MBS in a state corresponding to projector $\hat P$, after having applied the many-body propagator $\hat V$ followed by the local unitary $\hat U_2$. Contrasting with the left hand terms, the right hand side requires an ancilla qubit to measure. Preparing the ancilla in state $\ket{+} = \frac{1}{\sqrt{2}}(\ket{0} + \ket{1})$, we follow the same unitary protocol as in equation \eqref{fullevolution}, except in this case we use the single ancilla as the control for both unitaries $\hat U_1$ and $\hat U_2$. Written out, this produces the state
\beq
\label{phi}
\ket{\phi} = \frac{1}{\sqrt{2}}\l(\ket{0}_A\otimes \hat V \ket{\psi} + \ket{1}_A\otimes \hat U_2 \hat V \hat U_1 \ket{\psi} \r)\,.
\eeq
The real and imaginary parts of $\mean{\hat V^\dagger \hat P \hat U_2 \hat V \hat U_1}$ for the many-body state $\ket{\psi}$ are then the means of $\sigma_x \otimes \hat P$ and $\sigma_y \otimes P$ for the compound state $\ket{\phi}$,
\beq
\Mean{\hat V^\dagger \hat P \hat U_2 \hat V \hat U_1} = \bra{\phi} \sigma_x\otimes \hat P \ket{\phi} +  i \bra{\phi} \sigma_y\otimes \hat P \ket{\phi}\,.
\eeq
Preparing the state $\ket{\phi}$ of Equation \eqref{phi} requires the ancilla qubit to interact with both sites of the many-body system, but certain cases require only a single site interaction to measure $\mean{\hat V^\dagger \hat P \hat U_2 \hat V \hat U_1}$. This occurs when the post-selection projector takes a tensor product form, 
\beq
\label{tensor}
\hat P = \hat P_2\otimes \hat P_{1 E} \otimes \hat 1_{E'}\,.
\eeq
The product $\hat P \hat U_2$ can also be written in this way,
\beq
\hat P \hat U_2 = \l(\hat B_+ + i \hat B_-\r) \otimes  \hat P_{1 E} \otimes \hat 1_{E'}\,.
\eeq 
where we have written $\hat P_2 \hat U_2$ in terms of its Hermitian and anti-Hermitian parts, and as before we let $E'$ denote the (inaccessible) environmental degrees of freedom. Using this decomposition, it suffices to prepare the state
\beq
\label{phiprime}
\ket{\phi'} = \frac{1}{\sqrt{2}}\l(\ket{0}_A\otimes \hat V \ket{\psi} + \ket{1}_A\otimes \hat V \hat U_1 \ket{\psi} \r)\,,
\eeq
which requires only a controlled unitary between the ancilla qubit and site $1$ of the MBS. As before, the right hand side of \eqref{ineq} can then be written as a sum of observables,
\ba
\nonumber \Mean{\hat V^\dagger \hat P \hat U_2 \hat V \hat U_1} &= \bra{\phi'}\l(\sigma_x \otimes\hat B_+ - \sigma_y\otimes B_-  \r)  \ket{\phi'}\\
& + i \bra{\phi'}\l(\sigma_x \otimes\hat B_- + \sigma_y\otimes B_+  \r)  \ket{\phi'}\,.
\ea
We note that with identity \eqref{rho}, both of these procedures can be generalized to do complete state tomography. 

The ideas we have presented have potential applications in a variety of existing experimental setups. For example, current ion trap experiments have the potential to simulate spin models displaying long-range propagation of correlations\cite{Haffner2005a, Leibfried2005, Lin2009, Blatt2012,Allcock2013,Richerme2014}, making them amenable to the single ancilla protocol described above. The models studied have ground states that can be both prepared and measured through direct fluorescence spectroscopy following (if necessary) adiabatic passage. The protocol's ancilla qubit can take the form of either one of the ions present in the system or one of the global motional modes associated with the ion trap. A key requirement in these setups is the ability to individually address single ions in the experiment\cite{Nagerl1999,Johanning2009,Enferad2014,Piltz2014}. Alternatively, in optical lattice many-body simulators\cite{Bruder2005,Bakr2009,Greif2013,Sherson2010,Bloch2012} it is possible to use polarization of light as the ancilla qubit. Based on selection rules arising out of angular momentum conservation, the controlled unitary operation of the ancilla would correspond to a polarization-dependent interaction with a localized subsystem. Between interactions the light must be sent through a delay line (e.g., a Fabry-Perot cavity), so that correlations between spatially distant MBS sites are given enough time develop\cite{Lieb1972,Bravyi2006}.

\acknowledgements{The authors would like to thank Scott Glancy, Zhexuan Gong, Shelby Kimmel, Trey Porto, Christopher Monroe, Ken Brown, and Gerard Milburn for helpful input and discussion regarding this work. Support is provided by the NSF-funded Physical Frontier Center at the JQI.}

\bibliography{bibli}
\bibliographystyle{h-physrev}

\end{document}